\documentstyle[aps,prl,multicol]{revtex}
\begin{document}
\def\be{\begin{equation}}
\def\ee{\end{equation}}
\def\bc{\begin{center}}
\def\ec{\end{center}}
\def\bea{\begin{eqnarray}}
\def\eea{\end{eqnarray}}
\draft

\title{Non-perturbative renormalization of the KPZ growth dynamics}

\author{C. Castellano$^{(1)}$, M. Marsili$^{(2)}$
and L. Pietronero$^{(1,3)}$}
\address{$^{(1)}$ The ``Abdus Salam'' I.C.T.P.
Strada Costiera 11, I-34100 Trieste\\
$^{(2)}$
International School for Advanced Studies
(SISSA), via Beirut 2-4, Trieste I-34014\\
$^{(3)}$ Dipartimento di Fisica and INFM Unit,
University of Rome ``La Sapienza'', I-00185 Roma}

\maketitle 

\begin{abstract}
We introduce a nonperturbative renormalization approach which identifies
stable fixed points in any dimension for the Kardar-Parisi-Zhang dynamics
of rough surfaces.
The usual limitations of real space methods to deal with
anisotropic (self-affine) scaling are overcome with an indirect
functional renormalization.
The roughness exponent $\alpha$ is computed for dimensions
$d=1$ to 8 and it results to be in very good 
agreement with the available simulations.
No evidence is found for an upper critical dimension. 
We discuss how the present approach can be extended 
to other self-affine problems.
\end{abstract}

\begin{multicols}{2}
\narrowtext

Several models for the growth of rough surfaces appear to belong to the
universality class of the Kardar-Parisi-Zhang (KPZ) equation\cite{kpz,hz}:
\be
{\partial h(x,t) \over \partial t} = \nu \nabla^2 h + {\lambda \over 2}
(\nabla h)^2 + \eta(x,t).
\ee
Here the first term on the r.h.s. corresponds to surface tension,
the second describes lateral growth \cite{kpz} and $\eta$ is white noise.
The large scale properties of growing surfaces are described
by two exponents: the roughness exponent $\alpha$, defined by
the typical height fluctuation $\delta h\sim L^\alpha$ 
over a linear distance $L$, and the dynamic exponent $z$\cite{hz}.
For the KPZ equation, the exponents satisfy the
scaling relation $z+\alpha=2$ as a consequence of
Galileian invariance in the related Burgers equation \cite{fns}.
It is then sufficient to concentrate attention 
on $\alpha$.
In the usual field theory approach a perturbation theory is defined
with respect to the non-linear term ($\lambda$).
The corresponding renormalization group (RG) 
reveals \cite{hz} that the 
physics of the KPZ is related to a strong coupling fixed point, 
which is unaccessible by perturbative methods.
Except for $d=1$, where, thanks to a 
fluctuation-dissipation theorem, an analytic solution is possible,
the problem appears of strong coupling nature and
not much can be done from this theoretical perspective.

Extensive numerical simulations have been carried out for 
restricted solid on solid (RSOS) models\cite{ala}, for $d=1$ to $7$. 
These are discrete models which belong to the KPZ universality
class \cite{repinv} and are optimal for numerical 
simulations \cite{hz,ala}. These studies show a 
gradual decrease of the value of $\alpha$, without evidence for 
an upper critical dimension.
On the other hand, several analytical approaches\cite{hh} 
suggest an upper critical dimension $d_c=4$ above which one should 
have $\alpha=0$. 
% This discrepancy is currently very debated\cite{debate}.

In strong coupling situations, real space methods 
perform sometimes much better than continuum field theory.
This is particularly true for self-organized non-equilibrium systems
such as fractal growth\cite{dla}, sandpile models \cite{socrg}
and models with quenched disorder \cite{bsrg} where real
space methods give quite accurate results.
However, rough surfaces show anisotropic scaling properties
(self-affinity) and, in such a case, it is difficult to define 
the ``internal degrees of freedom'' for the block transformation 
in real space.
Here we overcome this basic problem by defining the
renormalization transformation in an indirect way that permits to 
integrate over the internal degrees of freedom without
the necessity of specifying them explicitly.
This approach is rather general and it should be useful also for 
other self-affine structures.

Our starting point is to introduce a parametrization of the 
dynamics at a generic scale. Then we devise a RG where
a scale transformation defines a flux in the parameter space 
of possible dynamics. The fixed point of the renormalization
transformation defines the {\em scale invariant dynamics} and
it describes the asymptotic scaling properties of the system.
The key point lies in the choice of a correct 
parametrization of the dynamics which 
captures the physics of the problem but is also 
simple enough to allow for explicit calculations. 
The main feature of the KPZ dynamics is lateral 
growth\cite{kpz,repinv}.
This allows us to concentrate on just one renormalization 
parameter which is the ratio between lateral and vertical growth 
rates.
At the simplest level the method is fully analytical while for 
accurate results or $d>1$ we resort to the Monte Carlo
method to compute the renormalization functions.
We obtain stable and attractive fixed points for any dimension from
$d=1$ to $8$. From them we compute the roughness exponent $\alpha$ 
which results to be in excellent agreement with the available 
simulations\cite{ala} up to $d=7$.
From the values of $\alpha$ and the stability of the fixed points
we conclude that there is no upper critical dimension, at
least up to $d=8$.

Let us consider a rough surface and define
a block transformation for its profile as indicated in 
Fig.~\ref{Fig1}. 
The coarse grained description is obtained 
by a partition of the $d+1$ dimensional space in cells
of lateral size $L_k=L_0 \cdot 2^k$ and thickness 
$h_k \simeq L_k^{\alpha}$. Each cell of the partition
is then declared to be full or empty according to some
majority rule.
With respect to the usual RG approaches one should note 
that the definition
of the coarse-graining transformation depends explicitly on the
exponent $\alpha$ which, on its turn, is just what 
we should compute in the end.
In addition the shape of the block variables changes with the scale
because, since in general $\alpha \neq 1$, the ratio $h_k/L_k$ will
depend on the scale $k$.
For these reasons the development of real space methods 
for self-affine problems has been extremely limited.

With this coarse grained description of the interface
in mind, let us now introduce the {\em effective} 
dynamics at the same level of coarse graining.
The dynamics at scale $k$ is defined in terms of 
transitions rates.
At the microscopic scale these correspond 
to the possible growth processes occurring along the 
surface, as specified by the model definition (for 
example in Eden growth vertical and lateral growth
have the same rates\cite{hz}).
At a generic scale $L_k$ we generalize this situation
in order to allow for proliferation in the RG.
We set the rate of the vertical growth nominally equal to $L_k$
and define a new parameter, $\lambda_k$, for the lateral growth.
The growth rate for a generic scale $k$ can then be expressed as:
\be
P[h(i)\to h(i)\!+\!1] \equiv L_k\!+\!\lambda_k \!
\sum_{j nn i}\max[0,h(j)-h(i)]
\label{rates}
\ee
The term $L_k$ is the contribution of the vertical growth and the
sum over neighbor block sites $j$ counts the contribution of lateral
growth.
We assume that $h(i)$ is expressed in units of $h_k$ 
and we neglect overhangs \cite{overh}.

We can now address the following question:
Given a description of the system at scale $k$, defined by
the parameters ($L_k$, $h_k$, $\lambda_k$)
what will be corresponding properties ($L_{k+1}$, $h_{k+1}$, 
$\lambda_{k+1}$)
at the scale $L_{k+1}= 2 L_k$?

A standard answer to this question, in the real space RG 
framework, is to partition a configuration at a given
scale into sub-configurations at a finer scale of coarse
graining (thereby integrating over degrees of freedom). 
This method is problematic because of the self-affine 
nature of the problem. 
A different method, which we introduce here, 
is to consider a system of size $L=L_{k+2}$ and 
its descriptions at two different scales, say $L/2=L_{k+1}$ 
and $L/4=L_k$. Imposing that the computation of the same 
quantity in the two different descriptions gives the same
result, leads to a RG equation relating
the dynamics at scale $k+1$ to that at scale $k$.
The natural quantity to consider in the case of surfaces is
the roughness which, for a system of size $L$, is given by
\be
W^2(L) = {1 \over L^d} \sum_{i} [h(i)-\bar h]^2
\ee
where $\bar h$ is the average height.
This has the nice property, in the perspective of the RG calculation,
that if one partitions the system in cells of size $b$
\be
W^2(L)=W^2(L/b)+W^2(b)
\label{wrg}
\ee
i.e. the roughness of the system is equal to the average of the
roughness in the individual cells, plus the roughness of the 
average height of the various cells.
Our strategy will then be to compute the roughness
$W^2_{k+2}=W^2(L_{k+2})$ over a linear size $L_{k+2}$ first 
using the description at scale $k+1$
and then at scale $k$. 

In order to do this we can first consider a system of two 
cells of size $L_{k+1}$ and then a system of four cells
at size $k$.
For simplicity we make the assumption of RSOS also for 
the coarse grained cells.
This is actually an approximation because, upon renormalization, a
dynamics which is of the restricted type ($|\Delta h| \leq 1$) at the
smallest scale can easily proliferate into an unrestricted one.
However with this restriction, the number of possible
surface configurations is very small and an analytic treatment
is possible. For the system of two sites we have only two 
configurations, that we classify by their height ($h_1$,$h_2$).
The first one is the flat one (1,1), while the second has one 
step (1,2).
With periodic boundary conditions and the definition 
(Eq. (\ref{rates})) of the dynamics at scale $k$, the transition
rates for the system of two configurations are
$P_{1\to2} = 2L_k$ and $P_{2\to1} = L_k + 2 \lambda_k$.
Indeed in $P_{1\to 2}$ only vertical growth is
possible (in both cells) whereas in configuration
$2$ only one site can grow and it has two lateral
contributions. 
The master equation for the probability $\varrho_i$ of 
configuration $i$ is
\be
{\partial \varrho_i \over \partial t} = 
\sum_{j} \varrho_j P_{j \to i}
-\varrho_i \sum_{j} P_{i \to j}
\label{master}
\ee 
and its stationary solution is $\varrho_2 = 1-\varrho_1 = 
2/(3+2 x_k)$, where $x_k=\lambda_k/L_k$.
Using eq. (\ref{wrg}), we can now compute the fluctuations at 
scale $k+1$
\be
W^2_{k+1} = W^2_{k} + W^2(2,x_k) = W^2_k + \varrho_2(x_k)
h_k^2/4.
\ee
Here $W^2_k$ is the fluctuation at scale $L_k$ which is present
in both configurations and $W^2(2,x_k)$ is the roughness 
corresponding to the fluctuation of the average height 
at scale $L_k$. Clearly the flat configuration does not
contribute and the roughness of the second configuration is
$h_k^2/4$.
We can relate $W^2_k$ to $h^2_k$ by observing that at scale 
$k$ the essential information we retain on the height distribution 
is only the discretization scale $h_k$. This is equivalent
to assuming that the distribution of the height fluctuations 
at this scale is a sum of two delta peaks
\be
P(\Delta h)= {1\over 2} \delta\left(\Delta h - {h_k\over 2}\right) +
             {1\over 2} \delta\left(\Delta h + {h_k\over 2}\right)
\ee
It follows that $W_k^2 = h_k^2/4$ and
\be
\frac{W^2_{k+1}}{W^2_k} \equiv F_1(x_k) = 1+
{2 \over 3+2 x_k}.
\label{Eq10}
\ee
This, at the moment, is a purely formal relation since $x_k$
is unknown. Clearly eq. (\ref{Eq10}) also relates
$W^2_{k+2}/W^2_{k+1}$ to $x_{k+1}$, and combining these two 
relations we get:
\be
\frac{W^2_{k+2}}{W^2_k} = 
\frac{W^2_{k+2}}{W^2_{k+1}}\cdot
\frac{W^2_{k+1}}{W^2_k}
 = F_1(x_k) F_1(x_{k+1}).
\label{wk1}
\ee

The roughness $W^2_{k+2}$ can also be computed directly
using the dynamics at scale $k$. For this
we need to consider a system of four sites, at scale $k$.
There are, always within 
the RSOS approximation and with periodic boundary conditions, 
six possible configurations of $(h_1,h_2,h_3,h_4)$.
The transition rates $P_{i\to j}$ are computed using 
eq. (\ref{rates}) and 
the stationary solution $\varrho_i$ of the master equation
(\ref{master}) is easily found. This allows one to compute
the roughness $W^2_{k+2}$ for the $4$ cells system
following exactly the same procedure used for the
$2$ cells system, obtaining
\be
\frac{W^2_{k+2}}{W^2_k}=F_2(x_k)=
1 + {51 + 86 x_k + 40 x_k^2 \over
47 + 106 x_k + 68 x_k^2 + 8 x_k^3}
\label{wk2}
\ee
Consistency of the two descriptions requires that
Eqs. (\ref{wk1}) and (\ref{wk2}) must yield the same
quantity. This leads to our basic RG transformation 
for $x_{k+1}$ as a function of $x_k$
\be
F_1(x_{k+1}) = F_2(x_k)/F_1(x_k).
\label{Eq14}
\ee
It can also be cast into the more conventional
form $x_{k+1}=R(x_k)$. A fixed point 
$x^* = R(x^*) \simeq 2.08779... $ exists and
we can compute the roughness exponent as
\be
\alpha = {1 \over 2} \log_2 \left({W^2_{k+1} \over W^2_k} \right)
= {1 \over 2} \log_2 F_1(x^*) 
\simeq 0.177352... 
\label{alpha}
\ee
The fixed point is also {\em attractive}. 
Close to the fixed point, we have $x_{k+1}-x^*=R(x_k)-x^*\simeq
R'(x^*)(x_k-x^*)$. If $|R'(x^*)|<1$, the distance 
$x_k-x^*\to 0$ as $k\to\infty$. We find $R'(x^*)=-0.03548...$ 
i.e. $x^*$ is strongly attractive.

The value of the exponent is 
clearly too small with respect to the exact
one ($\alpha= 1/2$). This is due to the RSOS approximation
which, on the other hand, allows for a fully analytical 
treatment. The point is that even if at the microscopic 
scale the dynamics of the surface is of restricted type 
(RSOS), the effective dynamics for the block variables at 
a generic scale $k$ defined by the renormalization
procedure can proliferate in a non restricted dynamics.
Allowing larger steps makes the problem 
analytically intractable. Still the calculation 
of the functions $F_1$ and $F_2$ can be done 
using a Monte Carlo method: For each value of the
parameter $x_k=\lambda_k/L_k$, the growth process 
in cells of sizes $b=2$ and $4$ described by eq. (\ref{rates}),
is numerically simulated. 
$F_1(x)$ and $F_2(x)$ are then computed as time averages 
in the stationary state from the roughness $W^2(b)$.
In this way one sees that, allowing for larger steps, 
the value of $\alpha$ approaches 
the exact value $\alpha=0.5$ with great accuracy.
Convergence is reached 
when steps $|\Delta h| \approx 8$ are allowed. 
Going further, i.e. to $16$, $\alpha$
does not change any more\cite{long}.

The Monte Carlo calculation of the functions $F_1(x)$ and $F_2(x)$
also allows to deal with higher dimensions $d>1$.
The function $F_1$ is then related to the roughness of a cell
of size $2^d$, while $F_2$ refers to $4^d$.
The method is exactly the same as for the one dimensional 
case.
The results for $d=1,\ldots 8$ are reported in Fig.~\ref{Fig2}.
Instead of $R(x)$, Fig.~\ref{Fig2} displays the function
$\tilde R(\alpha)$ where $\alpha(x)$ is given in the
second equality of eq. (\ref{alpha}). This method is graphically
convenient since in this way the exponents can be read from
the intersection of $\tilde R(\alpha)$ with the line $\alpha$
and the stability of the fixed point can be inferred from the 
slope $\tilde R'(\alpha^*)$ of the function at the fixed point.
We see that for $d=1,\ldots,8$ a fixed point with $\alpha>0$
exists and it is attractive (see inset).
The values of the exponent $\alpha$ are reported in table 
\ref{tabella}, together with the fixed point value $x^*(d)$
for $d=1,\ldots,8$. For comparison, table \ref{tabella} 
also reports the exponent $\alpha$
obtained in ref. \cite{ala} from numerical simulations up to 
$d=7$. The agreement between our RG results and the
simulations is rather impressive and it provides a strong
support to the validity of both methods. 

The extrapolation to $d \to \infty$ of $R'(x^*)$ suggests that the
fixed point never becomes unstable, leading to the conclusion that
no finite upper critical dimension exists for the KPZ dynamics
(note that $\alpha=0$ in our scheme would imply a stable
fixed point at $x^*=\infty$).
On the basis of these results, the arguments 
\cite{hh} leading to an upper critical 
dimension $d_c=4$ should probably be reconsidered. 
Furthermore our results suggests a behavior of the type
$\alpha(d)\sim 1/d$ and $x^*(d)\sim 2^d$ for large $d$. 
Indeed an expansion of our RG method for large 
dimensions, to be reported elsewhere\cite{long}, 
leads to such a behavior.
Finally we verified that our RG scheme is stable
with respect to the change of the cell sizes used.
Indeed a RG function $x_{k+1}=R(x_k)$ can also
be defined considering the functions 
$F_{\ell+1}(x)=W^2_{k+\ell+1}/W^2_k$, 
$F_{\ell}(x)=W^2_{k+\ell}/W^2_k$ and $F_1(x)$
(the case discussed above being $\ell=1$).
In $d=1$ we found essentially the same results,
whereas in larger dimensions we found the same
qualitative picture with a weak 
dependence of the exponent on $\ell$ \cite{long}.

In summary we have introduced a new method  
to renormalize self-affine dynamics in real space.
The key technical element is to effectively integrate 
over internal degrees of freedom in an indirect way 
{\em via} a consistency relation. This allows us to 
overcome the difficulty of defining a geometric
fine graining procedure for self-affine structures.
In some sense, the method generalizes 
the Fixed Scale Transformation \cite{dla} 
idea to self-affine structures. 
There are two key steps: first we introduce
a description of the system and the dynamics
at a generic scale $L_k$ in terms of the
parameters $h_k$ and $\lambda_k=x_k L_k$.
Secondly we derive the RG equation 
$x_{k+1}=R(x_k)$ with a new indirect method.
The attractive fixed point $x^*$
(self-organization) of the RG transformation 
identifies the {\em scale-invariant dynamics} and
allows  us to compute the exponent $\alpha$.

From the point of view of the approximations
involved in this method, we observe that the first
step is essentially a description tool and, as such,
it is in principle exact. The approximations arise
in the identification of the scale invariant 
dynamics, which can be made more and more accurate
with the introduction of additional proliferation parameters. 
This explains the striking accuracy of the present method 
as compared to usual real space RG methods for 
equilibrium phenomena.

Usually real space RG methods are at their best for 
low dimensional systems. 
The topological complexity of interactions indeed
increases with dimension thus requiring more and more 
drastic approximations for larger $d$. 
This is not the case here: Our method is exactly the same 
in any dimension and the computational complexity is 
overcome with Monte Carlo tecniques.

The nature of our approach is quite different from
the usual perturbation techniques for KPZ. Indeed
our parametrization of the dynamics does not
include the Edwards-Wilkinson equation ($\lambda=0$ in
eq. (1)). Rather the non-interacting limit of our theory 
is random deposition ($\lambda=\nu=0$).

Finally we observe that this general method can be
applied to a wide variety of other dynamic critical
phenomena. In this sense the present application to 
the KPZ problem is a very promising example.

We acknowledge interesting discussions with A. Maritan,
A. Stella, C. Tebaldi and A. Vespignani.

\begin{figure}
\caption{Coarse grained description at scale $L_k$ of a
growing interface.}
\label{Fig1}
\end{figure}

\begin{figure}
\caption{Renormalization group transformation
$\tilde R(\alpha)$ for $d=1,\ldots,8$. The intersections
with the line $\alpha$ yields the fixed point value 
$\alpha(x^*)$ of the exponent. The exponents
$\alpha$ and the slope $\tilde R'(\alpha)$
at the fixed point, are plotted in the inset as a function
of $d$. Since $|\tilde R'|<1$ we conclude that the fixed
points are stable.}
\label{Fig2}
\end{figure}

\end{multicols}
\widetext

\begin{table}
\begin{center}
\begin{tabular}{c|cccccccc}
$d$                & 1    & 2    & 3    & 4    & 5    & 6    & 7     & 8    \\ 
\hline 
$x^*$              & 0.735& 2.76 & 7.16 & 15.35& 31.8 & 62.5 & 125   & 240  \\
$\alpha_{\rm RG}$  & 0.502& 0.360& 0.284& 0.238& 0.205& 0.182& 0.162 & 0.150 \\
$\alpha_{\rm num}$ & 0.5  & 0.387& 0.305& 0.261& 0.193& 0.18 & 0.15  &  -   \\
\end{tabular}
\caption{Results of the RG approach compared with numerical simulations. 
The first line gives the value of the fixed point critical
parameter $x^*$. The fixed point is attractive and the corresponding
exponent $\alpha_{\rm RG}$ is listed in the second line. The third line
line reports the numerical results of ref. [4]. The excellent agreement
agreement between our RG result and numerical simulations provides a 
strong support for the validity of both, and it points to the absence an
upper critical dimension for this problem.}
\end{center}
\label{tabella}
\end{table}


\begin{thebibliography}{99}
\bibitem{kpz} 
M. Kardar, G. Parisi and Y. C. Zhang, 
Phys. Rev. Lett. {\bf 56}, 889 (1986).

\bibitem{hz} 
T. Halpin-Healy and Y.-C. Zhang, 1995, Phys. Rep. {\bf 254}.

\bibitem{fns} 
D. Forster, D. R. Nelson  and M. J. Stephen,
Phys. Rev. A {\bf 16}, 732 (1977).

\bibitem{ala} 
T. Ala-Nissila,  {\em et al}, 
%T. Hjelt, and J. M. Kosterlitz and O. Vem\"ol\"oinen,
J. Stat. Phys. {\bf 72}, 207 (1993);
L.-H. Tang, {\em et al},  
% B. M. Forrest and D. E. Wolf, 
Phys. Rev. A {\bf 45}, 7162 (1992).

\bibitem{repinv} 
M. Marsili, A. Maritan, F. Toigo and J. R. Banavar,
Rev. Mod. Phys. {\bf 68}, 963 (1996).

\bibitem{hh}
T. Halpin-Healy, Phys. Rev. A {\bf 42}, 711 (1990);
J-P. Bouchaud and M.E. Cates, Phys. Rev. E {\bf 47}, 1455 (1993);
M.A.  Moore {\em et al}, Phys. Rev. Lett. {\bf 74}, 4257 (1995);
M. L\"assig and H. Kinzelbach, Phys. Rev. Lett. {\bf 78}, 906 (1997).

%\bibitem{debate}
%T. Ala-Nissila, Phys. Rev. Lett. {\bf 80}, 887 (1998); J. M. Kim
%{\em ibid} 888, M. L\"assig and H. Kinzelbach, {\em ibid} 889.

\bibitem{dla} 
A. Erzan, L. Pietronero and A. Vespignani, 
Rev. Mod. Phys. {\bf 67}, 545 (1995); 
R. Cafiero, A. Vespignani and L. Pietronero 
Phys. Rev. Lett. {\bf 70}, 3939 (1993).

\bibitem{socrg} 
L. Pietronero, A. Vespignani and S. Zapperi
Phys. Rev. Lett. {\bf 72}, 1690 (1994); 
A. Vespignani, S. Zapperi and L. Pietronero,
Phys. Rev. E {\bf 51}, 1711 (1995).

\bibitem{bsrg}
M. Marsili, Europhys. Lett. {\bf 28}, 385 (1994);
A. Gabrielli, R. Cafiero, M. Marsili and L. Pietronero,
Europhys. Lett. {\bf 38}, 491 (1997).

\bibitem{overh} 
Overhangs have been proved to be irrelevant in M. Marsili and 
Y.-C. Zhang, Phys. Rev. E, to appear (1998).

\bibitem{long}
C. Castellano, M. Marsili and L. Pietronero, preprint (1998).

\end{thebibliography}
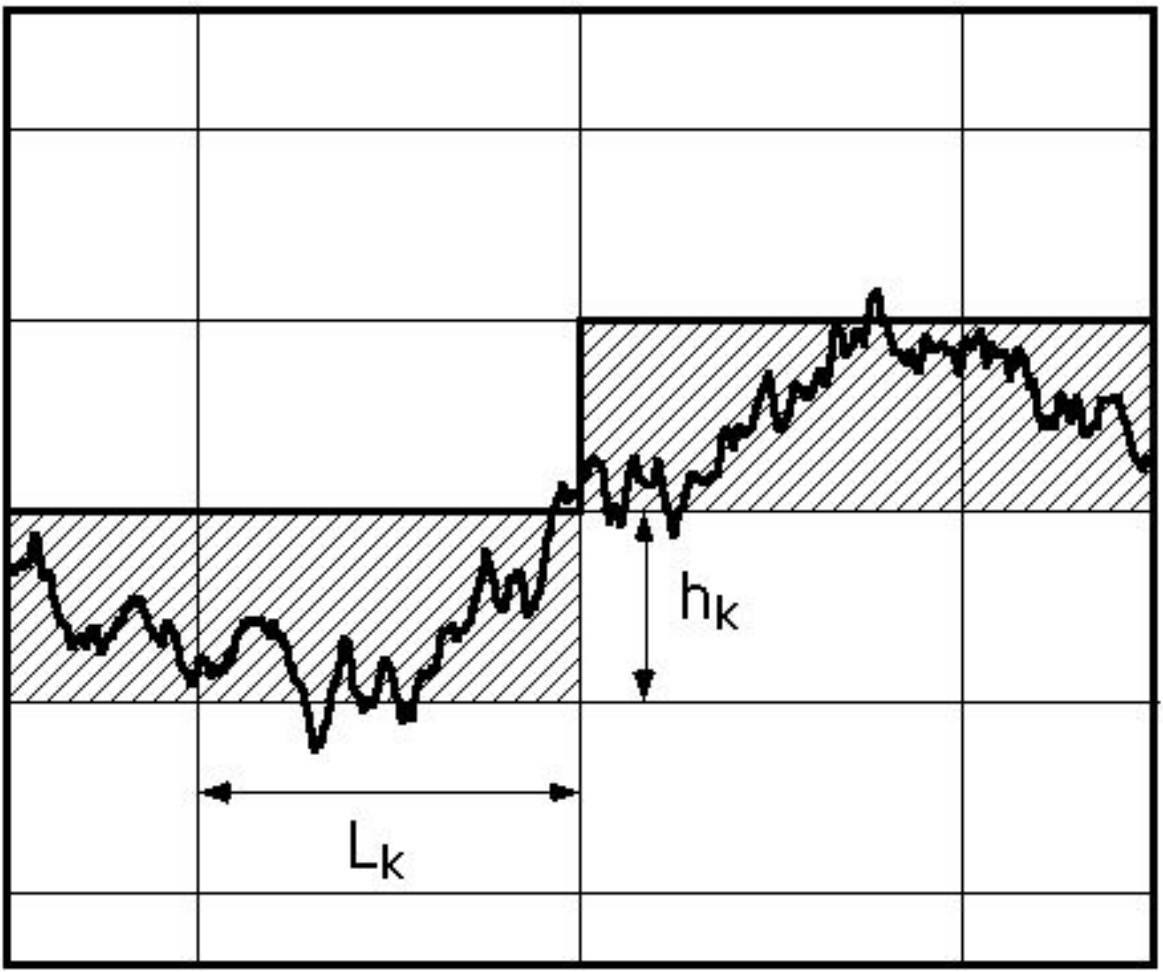
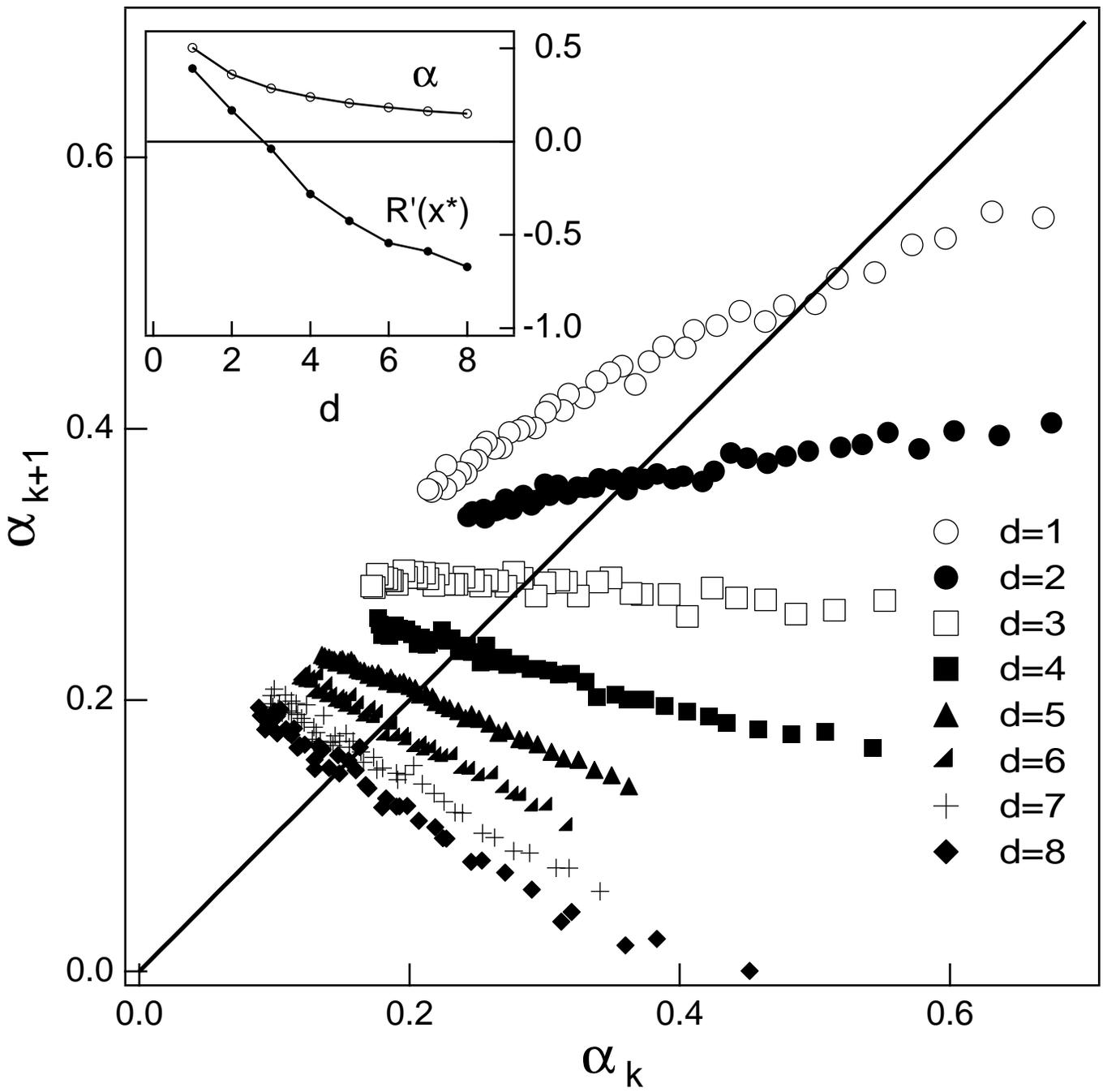
\end{document}